\documentstyle[aps,12pt,cite]{revtex}
\input epsf

\begin{document}

\title {Analysis of nonperturbative fluctuations in a triple-well potential}
\author { J. Casahorr\'an \footnote{email:javierc@posta.unizar.es}}
\address{Departamento de F\'{\i}sica Te\'orica, \\
Universidad de Zaragoza, E-50009 Zaragoza, Spain}

\maketitle

\begin{abstract}

We consider  the quantum tunneling phenomenon in a well-behaved triple-well
potential. As required by the semiclassical approximation we take into
account the quadratic fluctuations over the instanton which represents as
usual the localised finite-action solution of the euclidean equation of
motion. The determinants of the quadratic differential operators at issue are
evaluated by means of the Gelfang-Yaglom method. In doing so the explicit
computation of the conventional
ratio of determinants  takes  as reference the harmonic
oscillator whose frequency is the
average of the individual frequencies derived from
the non-equivalent minima of the potential. Eventually the physical
effects of the multi-instanton configurations are included in this approach.
As a matter of fact we obtain information about the energies of the
ground-state
and the two first excited levels of the discrete spectrum at issue.

\end{abstract}

\vfill \eject

The semiclassical analysis of the tunneling phenomenon, ranging from
periodic-potentials in quantum mechanics to Yang-Mills models in field
theories, has been performed in a systematic way by means of the instanton
method. To be precise, the instanton represents itself a localised
finite-action solution of the euclidean equation of motion. In principle
one finds the classical configuration and subsequently evaluate the
quadratic fluctuations. The functional integration at issue is solved in
terms of the gaussian scheme except for the zero-modes which appear by
virtue of the translational invariance of the model. Once the collective
coordinates technique is incorporated the gaussian integration takes into
account the directions orthogonal to the zero-modes. As a functional
determinant includes an infinite product of eigenvalues, the result
should be a highly divergent expression. However we can regularize the
fluctuation factors by means of the ratio of determinants. \par

Let us describe in brief the instanton calculus for the one-dimensional
particle as can be found for instance in \cite{kl}. Unless otherwise
noted we assume that our particle moves under the action of a confining
potential $V(x)$ which yields a pure discrete spectrum of energy eigenvalues.
In addition we choose the origin of the energy so that the minima of the
potential satisfy $V(x) = 0$.
If the particle is located at the initial time $t_i = -T/2$ at the point
$x_i$ while one finds it when $t_f = T/2$ at the point $x_f$, the
functional version of the non-relativistic quantum mechanics allows us to
express the transition amplitude in terms of a sum over all paths joining
the world points with coordinates $(-T/2, x_i)$ and $(T/2, x_f)$.
Making the change $t \rightarrow - i \tau$, known in the
literature as the Wick rotation, the euclidean formulation of the
path-integral reads

\begin{equation}
<x_f\vert \exp(- H T) \vert x_i> = N(T) \int [dx]  \ \exp - S_e[x(\tau)]
\label{eq:1}
\end{equation}

\noindent where $H$ represents the hamiltonian of the model, the
factor $N(T)$ serves to normalize the amplitude conveniently while
$[dx]$ indicates the integration over all functions which fulfil the
boundary conditions at issue. In addition we resort to the euclidean
action $S_e$, i.e.

\begin{equation}
S_e =  \int_{-T/2}^{T/2} \left[  {{1} \over {2}}
\left( {{dx} \over {d\tau}} \right)^2 + V(x) \right]  \ d\tau
\label{eq:2}
\end{equation}

\noindent whenever the mass of the particle is set equal to unity for
notational simplicity. This semiclassical approximation  takes for
granted the existence of a $x_c(\tau)$ configuration which represents
a stationary point of the euclidean action. Next we perform the expansion
of a generic $x(\tau)$ according to

\begin{equation}
x(\tau) = x_c(\tau) + \sum_j c_j \  x_j(\tau)
\label{eq:3}
\end{equation}

\noindent where $x_j(\tau)$ stand for a complete set of orthonormal functions

\begin{equation}
\int_{-T/2}^{T/2} x_j(\tau) \  x_k(\tau) \ d\tau = \delta_{jk}
\label{eq:4}
\end{equation}

\noindent  vanishing at our boundary, i.e. $x_j(\pm T/2) = 0$. As a matter
of fact the eigenfunctions $x_j(\tau)$ (with eigenvalues $\epsilon_j$)
appear associated with the so-called stability equation given by

\begin{equation}
-  {{d^2 x_j(\tau)} \over {d\tau^2}}  +
 V^{\prime \prime}[x_c(\tau)] x_j(\tau) = \epsilon_j x_j(\tau)
\label{eq:5}
\end{equation}

In doing so the euclidean transition amplitude reduces itself to

\begin{equation}
<x_f\vert \exp(- H T) \vert x_i> = N(T) \exp(- S_{eo}) \ \prod_j
\epsilon_{j}^{-1/2}
\label{eq:6}
\end{equation}

\noindent where $S_{eo}$ represents
the classical action associated with the configuration $x_c(\tau)$ while the
product of eigenvalues is usually written as

\begin{equation}
\prod_j \epsilon_{j}^{-1/2} = \left\{ Det \left[
-  {{d^2} \over {d\tau^2}}  +
 V^{\prime \prime}[x_c(\tau)]  \right] \right\}^{-1/2}
\label{eq:7}
\end{equation}

\noindent according to a notation which obviously has its origin in
the finite-dimensional case. To fix the factor $N(T)$ we resort to a
well-known problem where

\begin{equation}
V(x) = {{\nu^2 } \over {2}} \  x^2
\label{eq:8}
\end{equation}

\noindent so that $V^{\prime \prime} (x = 0 ) = \nu^2$. As corresponds
to the harmonic oscillator the relevant amplitude is given by

\begin{equation}
<x_f = 0\vert \exp(- H_{ho} T) \vert x_i = 0> = N(T) \left\{ Det \left[
-  {{d^2} \over {d\tau^2}}  +
 \nu^2  \right]\right\}^{-1/2}
\label{eq:9}
\end{equation}

Next the evaluation of (\ref{eq:9}) is possible according to the general
method exposed in \cite{ra}. To sum up

\begin{equation}
<x_f = 0\vert \exp(- H_{ho} T) \vert x_i = 0> = \left({{\nu} \over
{\pi}}\right)^{1/2} \
\left(2 \sinh \nu T \right)^{-1/2}
\label{eq:10}
\end{equation}

In the following we consider the triple-well potential $V(x)$ given by

\begin{equation}
V(x) = {{\omega^2 } \over {8}} x^2 (x^2 - 1)^2
\label{eq:11}
\end{equation}

From a classical point of
view we find three minima located at $x_{-} = - 1$, $x_{+} =  1$ and
$x_{o} = 0$. Two of them, namely $x_{-}$ and $x_{+}$, are equivalent since
can be connected by means of the discrete symmetry $x \rightarrow - x$
at issue. However the third minimum $x_{o}$ is invariant under the action
of such a symmetry. In other words, the central vacuum is not identical
with the others located at both sides.
When considering the limit  $\omega^2 \gg 1$ the energy
barriers are high enough to decompose the system into a sum of
independent harmonic oscillators. However the existence of finite barriers
between the different wells of the potential yields a relevant
tunneling phenomenon so that ultimately the symmetry $x \rightarrow - x$ is
not spontaneously broken at quantum level and the expectation value of the
coordinate $x$ computed for the ground-state is zero as corresponds to the
even character of the potential $V(x)$. The question we wish to address now
is the explicit description of the tunneling in the euclidean version of the
path-integral.
As regards the one-instanton amplitude
we evaluate the transition amplitude between
$x_{o} = 0$ and $x_{+} =  1$. In doing so we need a classical configuration
with  $x_{i} = 0$ at $t_{i} = -T/2$ while
$x_{f} =  1$ when $t_{f} = T/2$.
It is customarily assumed that $T \rightarrow \infty$ mainly because
the explicit solution of the problem is much more complicated for finite $T$
(more on this point later). \par

As a matter of fact we can find the explicit form of the instanton
$x_{c}(\tau)$
just by integration of a first-order differential equation, i.e.

\begin{equation}
 {{dx_{c}} \over {d\tau}} = \pm \sqrt{ 2 V(x_{c})}
\label{eq:12}
\end{equation}

\noindent where we recognize the quantum mechanical version of the Bogomol'nyi
condition \cite{rj}.
Now we solve (\ref{eq:12}) by a simple quadrature so that

\begin{equation}
x_c(\tau) = \sqrt {[1 + \tanh \omega (\tau - \tau_c) ] / 2}
\label{eq:13}
\end{equation}

\noindent where the parameter $\tau_c$ indicates the point at which the
instanton
makes the jump.
As expected equivalent solutions are  obtained by means of the
transformation $\tau \rightarrow - \tau$ and $x_c(\tau) \rightarrow -
x_c(\tau)$ so
that adjoint  minima of the potential can be connected by means of a
topological
solution. In addition we have that $S_{eo} = \omega/4$. On the other hand we
need classical configurations for which $x_{-} = 0$ and $x_{+} = 1$ at large
but finite values $\tau = \pm T/2$. However the explicit form of the instantons
that appear in the literature corresponds to infinite euclidean time.
Fortunately
the difference is so small that can be ignored mainly because we are interested
in the limit $T \rightarrow \infty$. In any case our description of the
one-instanton amplitude between $x_{-} = 0$ and $x_{+} = 1$ takes over

\begin{eqnarray*}
<x_f = 1\vert \exp(- H T) \vert x_i = 0> = N(T) \left\{ Det \left[
- {{d^2} \over {d\tau^2}}  +
 \nu^2  \right]\right\}^{-1/2}
\end{eqnarray*}
\begin{equation}
\left\{{{Det \left[- (d^2/d\tau^2) + V^{\prime \prime}[x_c(\tau)] \right]}
\over
{Det \left[- (d^2/d\tau^2) + \nu^2 \right]}}\right\}^{- 1/2} \
\exp(-S_{eo})
\label{eq:14}
\end{equation}

\noindent where we have multiplied and divided by the determinant associated
with the harmonic oscillator of frequency $\nu$. As regards the determinant
built over the instanton itself we find a zero-mode $x_{o}(\tau)$ which could
jeopardize the computation process.
However this eigenvalue $\epsilon_o = 0$ comes by no surprise
since it reflects the translational invariance of the system.
 As a matter of fact one can discover the
existence of a zero-mode starting from (\ref{eq:13}). Including the adequate
normalization one can check that

\begin{equation}
x_{o}(\tau) = {{1} \over {\sqrt{S_{eo}}}} {{dx_{c}} \over {d\tau}}
\label{eq:15}
\end{equation}

\noindent is just the solution of (\ref{eq:5}) with $\epsilon_o = 0$.
The way out of this apparent cul-de-sac
is simple. The integration over $c_{o}$ (see (\ref{eq:3})) becomes equivalent
to the integration over the center of the instanton $\tau_{c}$. To fix the
jacobian of the transformation involved we take a first change such that

\begin{equation}
\Delta x(\tau) = x_{o}(\tau) \  \Delta c_{o}
\label{eq:16}
\end{equation}

According to the general expression written in (\ref{eq:3}) we find that under
a shift $\Delta \tau_{c}$ the effect should be

\begin{equation}
\Delta x(\tau) = - \sqrt{S_{eo}} \ x_{o}(\tau) \Delta \tau_{c}
\label{eq:17}
\end{equation}

Now the identification between (\ref{eq:16}) and (\ref{eq:17}) yields

\begin{equation}
d c_{o} = \sqrt{S_{eo}} \  d\tau_{c}
\label{eq:18}
\end{equation}

\noindent where the minus sign disappears since what matters is the modulus
of the jacobian at issue. In doing so we have that

\begin{eqnarray*}
\left\{{{Det \left[- (d^2/d\tau^2) + V^{\prime \prime}[x_c(\tau)] \right]}
\over
{Det \left[- (d^2/d\tau^2) + \nu^2 \right]}}\right\}^{- 1/2} =
 \end{eqnarray*}
\begin{equation}
\left\{{{Det^{\prime} \left[- (d^2/d\tau^2) + V^{\prime \prime}[x_c(\tau)]
\right]} \over
{Det \left[- (d^2/d\tau^2) + \nu^2 \right]}}\right\}^{- 1/2} \
\sqrt{{{S_{eo}} \over {2 \pi}}} \ d\tau_{c}
\label{eq:19}
\end{equation}

\noindent where $Det^{\prime}$ stands for the so-called reduced determinant
once the zero-mode has been removed. In order to make an explicit computation
of the quotient of determinants we resort to the well-grounded Gelfand-Yaglom
method so that only the knowledge of the large-$\tau$ behaviour of the
classical solution $x_{c}(\tau)$ is necessary \cite{gy}. Being $\hat{O}$ and
$\hat{P}$ a couple of se\-cond order differential operators, whose
eigenfunctions
vanish at the boundary, the quotient of determinants is given in terms of the
respective zero-energy solutions $f_{o}(\tau)$ and $g_{o}(\tau)$ according to

\begin{equation}
{{Det \hat{O}} \over {Det \hat{P}}} = {{f_{o}(T/2)} \over {g_{o}(T/2)}}
\label{eq:20}
\end{equation}

\noindent whenever the eigenfunctions at issue satisfy the initial conditions

\begin{equation}
f_{o}(-T/2) = g_{o}(-T/2) = 0, \ \ \ {{df_{o}} \over {d\tau}} (-T/2) =
{{dg_{o}} \over {d\tau}} (-T/2) = 1
\label{eq:21}
\end{equation}

As regards the zero-mode $g_{o}(\tau)$ associated with the harmonic oscillator
of frequency $\nu$ we find

\begin{equation}
g_{o}(\tau) = {{1} \over {\nu}} \  \sinh [\nu (\tau + T/2)]
\label{eq:22}
\end{equation}

\noindent so that now we need the explicit form of the solution
$f_{o}(\tau)$ which
corresponds to the topological configuration written in (\ref{eq:13}).
Starting from the $x_{o}(\tau)$ zero-mode we can construct a second solution
$y_{o}(\tau)$ given by

\begin{equation}
y_{o}(\tau) = x_{o}(\tau) \ \int_{0}^{\tau} {{ds} \over {x_{o}^{2}(s)}}
\label{eq:23}
\end{equation}

Accordingly we may summarize the asymptotic behaviour of $x_{o}(\tau)$ and
$y_{o}(\tau)$ as follows

\begin{equation}
x_{o}(\tau) \sim \left\{ \matrix{C \exp (- 2 \omega \tau) \ \ \ if \ \
\tau \rightarrow \infty \cr D \exp ( \omega \tau) \ \ \ \  if \ \
\tau \rightarrow - \infty \cr} \right.
\label{eq:24}
\end{equation}

\begin{equation}
y_{o}(\tau) \sim \left\{ \matrix{ \exp ( 2 \omega \tau)/4 \omega C \ \ \ if \ \
\tau \rightarrow \infty \cr - \exp (- \omega \tau)/2 \omega D \ \ \ \  if \ \
\tau \rightarrow - \infty \cr} \right.
\label{eq:25}
\end{equation}

\noindent where the constants $C$ and $D$ derive from the explicit form of
the derivative of (\ref{eq:13}). We proceed to investigate the particular
solution $f_{o}(\tau)$ which is the one we are really interested in. Starting
from the linear combination of $x_{o}(\tau)$ and $y_{o}(\tau)$ given by

\begin{equation}
f_{o}(\tau) = A x_{o}(\tau) + B y_{o}(\tau)
\label{eq:26}
\end{equation}

\noindent the incorporation of the initial conditions at issue leads us to

\begin{equation}
f_{o}(\tau) = x_{o}(- T/2) y_{o}(\tau) - y_{o}(- T/2) x_{o}(\tau)
\label{eq:27}
\end{equation}

From this expression, which is exact, we can extract the asymptotic behaviour
of $f_{o}(\tau)$, i.e.

\begin{equation}
f_{o}(T/2) \sim {{D} \over {4 \omega C}} \ exp(\omega T/2) \ \ \ \  if  \ \
T \rightarrow \infty
\label{eq:28}
\end{equation}

Next we need to take into account the lowest eigenvalue of the stability
equation to obtain the right value of the quotient of determinants. From
a physical point of view we can explain the situation as follows: the
derivative of the topological solution does not quite satisfy the boundary
conditions for the interval $(- T/2,T/2)$. When enforcing such a behaviour,
the eigenstate is compressed and the energy shifted slightly upwards. In doing
so the zero-mode $x_{o}(\tau)$ is substituted for the $f_{\lambda}(\tau)$ which
corresponds to

\begin{equation}
- {{d^2 f_{\lambda}(\tau)} \over {d\tau^2}}  +
 V^{\prime \prime}[x_c(\tau)] f_{\lambda}(\tau) = \lambda f_{\lambda}(\tau)
\label{eq:29}
\end{equation}

\noindent whenever

\begin{equation}
f_{\lambda}(- T/2) = f_{\lambda}(T/2) = 0
\label{eq:30}
\end{equation}

To lowest order in perturbation theory we get

\begin{equation}
f_{\lambda}(\tau) \sim  f_{o}(\tau) + \left.\lambda \ {{df_{\lambda}}
\over {d\lambda}} \right |_{\lambda = 0}
\label{eq:31}
\end{equation}

\noindent so that ultimately

\begin{equation}
f_{\lambda}(\tau) =  f_{o}(\tau) + \lambda \int_{-T/2}^{\tau}
[x_{o}(\tau) y_{o}(s) - y_{o}(\tau) x_{o}(s)] \ f_{o}(s) \ ds
\label{eq:32}
\end{equation}

The asymptotic behaviour of $f^{o}(\tau)$, $x_{o}(\tau)$ and $y_{o}(\tau)$,
together with the condition $f_{\lambda}(T/2) = 0$ allows us to find the
lowest eigenvalue $\lambda$, i.e.

\begin{equation}
\lambda = 2 \omega D \exp(- \omega T)
\label{eq:33}
\end{equation}

In such a case the Gelfand-Yaglom method provides us with the final expression
of the quotient of determinants whenever we choose for the frequency $\nu$ of
the harmonic oscillator of reference the average of the frequencies of the
central and lateral wells. In other words $\nu = 3 \omega /2$. Notice the
difference with the well-grounded double-well model where the two minima of
the potential are equivalent so that the aforementioned average is not
necessary. Going back to the explicit form of $x_{c}(\tau)$ (see (\ref{eq:13}))
we find $C = D = \omega/\sqrt{S_{eo}}$. Armed with this information we can
write
the one-instanton amplitude between the points $x_{i} = 0$ and $x_{f} = 1$ in
terms of

\begin{eqnarray*}
<x_f = 1\vert \exp(- H T) \vert x_i = 0> =
\end{eqnarray*}
\begin{equation}
 \left({{3 \omega} \over {2 \pi}}\right)^{1/2} \
\left(2 \sinh 3 \omega T/2 \right)^{-1/2} \  \sqrt{S_{eo}} \
\sqrt{{{4} \over {3 \pi}}} \
\exp(-S_{eo}) \  \omega \  d\tau_{c}
\label{eq:34}
\end{equation}

Apart from the first factor, which represents the contribution of the
harmonic oscillator of reference, we get a transition amplitude just
depending on the point $\tau_{c}$ at which the instanton makes
precisely the jump. \par

Although all the above calculations were carried out over a single instanton,
it remains to identify the  contributions which take into account the
effect of a string of widely separated
instantons and antiinstantons along the $\tau$ axis.
It is customarily assumed that these combinations of topological solutions
represent no strong deviations of the trajectories just derived from the
euclidean equation of motion without any kind of approximation. We shall
compute
the functional integral by summing over all such configurations, with
$k$ instantons and antiinstantons centered at points
$\tau_1,...,\tau_k$ whenever

\begin{equation}
-{{T} \over {2}} < \tau_1 < ... < \tau_k < {{T} \over {2}}
\label{eq:35}
\end{equation}

Being narrow enough the regions where the instantons (antiinstantons) make the
jump, the action of the
proposed path is almost extremal. We can carry things further and
assume that the action of the string of instantons and antiinstantons is
given by the sum of the $k$ individual actions. This scheme is well-known
in the literature where it appears with the name of dilute gas approximation
\cite{sh}.
In addition the translational degrees of freedom of the  separated $k$
instantons and antiinstantons yield an integral of the form

\begin{equation}
\int_{-T/2}^{T/2} \omega d\tau_j \
\int_{-T/2}^{\tau_k} \omega d\tau_{k - 1} ...
\int_{-T/2}^{\tau_2} \omega d\tau_1 = {{(\omega T)^k} \over {k!}}
\label{eq:36}
\end{equation}

As regards the quadratic fluctuations around the $k$ topological solutions
we have
now that the single ratio of determinants transforms into \cite{co}

\begin{eqnarray*}
\left({{3 \omega} \over {2 \pi}}\right)^{1/2} \
\left(2 \sinh 3 \omega T/2 \right)^{-1/2} \
\left\{{{Det^{\prime} \left[- (d^2/d\tau^2) + V^{\prime \prime}[x_c(\tau)]
\right]} \over
{Det \left[- (d^2/d\tau^2) + 9 \omega^2/4 \right]}}\right\}^{- 1/2}
\longrightarrow
\end{eqnarray*}
\begin{equation}
\left({{3 \omega} \over {2 \pi}}  \right)^{1/2} \exp(- 3 \omega T/4) \
\left[\left\{{{Det^{\prime} \left[- (d^2/d\tau^2) + V^{\prime
\prime}[x_c(\tau)] \right]} \over
{Det \left[- (d^2/d\tau^2) + 9 \omega^2/4 \right]}}\right\}^{- 1/2}\right]^k
\label{eq:37}
\end{equation}

\noindent according to the limit of the factor associated with the harmonic
oscillator when $T$ is large. In addition,
when going to the dilute-gas approximation we must consider a set of instantons
and anti-instantons so that each topological configuration starts where its
predecessor ends. For the amplitude at issue the total number $k$ of instantons
plus anti-instantons must be odd. On the other hand, the combinatorial
factor $F$ associated with the number of possible configurations corresponds to
$F = 2^{(k-1)/2}$ since the closed paths starting and coming back to the point
$x_{o} = 0$ require in a systematic way instanton-anti-instanton pairs.
Notice the difference with the double-well potential where the instantons
strictly alternate with the anti-instantons since the problem has only two
minima so that the combinatorial factor is not necessary at all.
Next we can write the
complete transition amplitudes for the triple-well potential so that

\begin{equation}
<x_f = 1\vert \exp(- H T) \vert x_i = 0> =
\left({{3 \omega} \over {4 \pi}}  \right)^{1/2} \exp(- 3 \omega T/4)
\sum_{j=0}^{\infty} {{(\omega T d)^{2j+1}} \over {(2j+1)!}}
\label{eq:38}
\end{equation}

\noindent where $d$ stands for the so-called instanton density, i.e.

\begin{equation}
d = \sqrt{{{8} \over {3 \pi}}} \  \sqrt{S_{eo}} \  \exp(-S_{eo})
\label{eq:39}
\end{equation}

In summary

\begin{equation}
<x_f = 1\vert \exp(- H T) \vert x_i = 0> =
\left({{3 \omega} \over {4 \pi}}  \right)^{1/2} \exp(-3 \omega T/4) \
\sinh (\omega T d)
\label{eq:40}
\end{equation}

Going back to the formal euclidean transition amplitude written in(\ref{eq:1}),
the insertion of the pure discrete spectrum of energy eigenfunctions, namely

\begin{equation}
H \vert n > = E_n \vert n >
\label{eq:41}
\end{equation}

\noindent allows us to write that

\begin{equation}
<x_{f} = 1\vert \exp(- H T) \vert x_{i} = 0> = \sum_{n}
\exp(- E_n T) <x_{f} = 1 \vert n >
<n\vert x_{i} = 0 >
\label{eq:42}
\end{equation}

Let us denote by $E_{o}$, $E_{1}$ and $E_{2}$ the energies of the ground-state
and the two first excited levels of our problem. As the triple-well potential
we are dealing with is even we know that $<1\vert x_{i} = 0 >$ vanishes so that
the limit $T \rightarrow \infty$ in (\ref{eq:40}) provides us with the
following
energy eigenvalues

\begin{equation}
E_{o} = {{3 \omega} \over {4}} - \omega d, \ \ \ \
E_{1} = {{3 \omega} \over {4}}, \ \ \ \
E_{2} = {{3 \omega} \over {4}} + \omega d
\label{eq:43}
\end{equation}

In other words, the tunneling phenomenon transfers along the whole real axis
the gaussian wave functions constructed on the points $x_{-} = - 1$,
$x_{+} =  1$ and $x_{o} = 0$. As expected the splitting term results
proportional to the barrier-penetration factor. The average of the harmonic
frequencies over the non-equivalent minima of the potential serves as the
central position for the splitting.

\vfill \eject


\begin{thebibliography}{99}

\bibitem{kl}
{H. Kleinert,} {\it Paths Integrals in Quantum Mechanics, Statistics and
Polymer Physics}.
Singapore: World Scientific (1990).
\bibitem{ra}
{P. Ramond,} {\it Field Theory: A Modern Primer}.
New York: Addison-Wesley Publishing Company (1992).
\bibitem{rj}
{R. Rajaraman,} {\it Solitons and Instantons}.
Amsterdam: North-Holland (1982).
\bibitem{gy}
{I.M. Gelfand and A.M. Yaglom,} Journ. Math. Phys. {\bf 1} (1960) 48.
\bibitem{sh}
{M.A. Shifman,} {\it ITEP Lectures on Particle Physics and Field Theory}.
Singapore: World Scientific (1999).
\bibitem{co}
{S. Coleman,} {\it Uses of Instantons} in {\it The Whys of Subnuclear Physics}.
Ed. A. Zichichi.
New York: Plenum Press (1979).

\end{thebibliography}
\end{document}